\documentclass[epsf,prb,twocolumn,showpacs]{revtex4-1}
\usepackage[pdftex]{graphicx}
\usepackage{dcolumn}
\usepackage{bm}
\usepackage{epsfig}
\usepackage{latexsym}
\usepackage{amsmath}
\usepackage{amsfonts}
\usepackage{amssymb}
\usepackage{color}
\usepackage{array}

\newcommand{\<}{\langle}
\newcommand{\e}{\varepsilon}
\newcommand{\up}{\uparrow}
\newcommand{\down}{\downarrow}
\renewcommand{\>}{\rangle}
\renewcommand{\(}{\left(}
\renewcommand{\)}{\right)}
\renewcommand{\[}{\left[}
\renewcommand{\]}{\right]}
\renewcommand{\v}[1]{\mathbf{#1}} 

\newcommand{\tv}{\tau_v}
\newcommand{\tvb}{\tau_{\bar{v}}}
\newcommand{\T}{\mathcal{T}}

\newcommand{\be}{\begin{equation}}
\newcommand{\ee}{\end{equation}}
\newcommand{\bea}{\begin{eqnarray}}
\newcommand{\eea}{\end{eqnarray}}

\begin{document}

\title{Gapped Symmetry Preserving Surface-State for the Electron Topological Insulator}
\author{Chong Wang, Andrew C. Potter, and T. Senthil}
\affiliation{Department of Physics, Massachusetts Institute of Technology,
Cambridge, MA 02139, USA}
\date{\today}
\begin{abstract}
It is well known that the 3D electronic topological insulator (TI) with charge-conservation and time-reversal symmetry cannot have a trivial insulating surface that preserves symmetry.  It is often implicitly assumed that if the TI surface preserves both symmetries then it must be gapless.  Here we show that it is possible for the TI surface to be both gapped and symmetry-preserving, at the expense of having surface-topological order.  In contrast to analogous bosonic topological insulators, this symmetric surface topological order is intrinsically non-Abelian.    We show that the surface-topological order provides a complete non-perturbative definition of the electron TI that transcends a free-particle band-structure picture, and could provide a useful perspective for studying strongly correlated topological Mott insulators. 
\end{abstract}
\maketitle

\section{Introduction}
In the last decade, dramatic progress has been made in understanding the topological properties of non-fractional electronic insulators\cite{Reviews}. While the original theoretical constructions were framed in terms of band structures for non-interacting electrons, attention has recently turned towards the interplay of strong correlation and topological insulation. It is now appreciated that the electron topological insulator is part of a larger class of quantum phases of matter known as Symmetry Protected Topological (SPT) phases\cite{chencoho2011}.  SPT phases are defined as states with a bulk gap and bulk excitations that do not carry fractional quantum numbers or exotic statistics, but which nevertheless have non-trivial surface states protected by symmetry. A well-known example is the Haldane spin chain in one dimension\cite{HaldaneChain}. 
 
It is important to recognize the difference between SPT states and more exotic states like those exhibiting the fractional quantum Hall effect which have what is known as ``intrinsic" topological order\cite{Wenbook}.  Topologically ordered states of matter have ground state degeneracies on topologically non-trivial manifolds  and, in the presence of symmetry, may have excitations with fractional quantum numbers. SPT phases on the other hand have a unique ground state and no fractional quantum numbers in the bulk.  Thus SPT phases do not have topological order. 
 
The Fu-Kane-Mele electronic TI (eTI)\cite{FKM} is the first known 3D example of an SPT phase. Its non-trivial surface states are protected by bulk time-reversal symmetry (TRS) and charge conservation ($U(1)_C$) symmetry.  If either of these symmetries is broken in the bulk, the eTI can be smoothly deformed into a trivial insulator.
It is, by now, well known that the surface can either be 1) a gapless, symmetry-preserving state, or 2) a gapped state that breaks one (or both) of TRS and $U(1)_C$.  
For some time, it was implicitly assumed that these options exhausted the possible surface phases.  Indeed these are the only possibilities accessible in a weakly interacting description of the surface. However in the presence of strong correlations other options for the surface may become available. In particular, we will show that it is possible for the eTI surface to be both fully gapped and preserve all symmetries. The price to pay for having a gapped and symmetric surface is that the surface develops intrinsic topological order (even though the bulk does not). We describe this surface topologically ordered state of the eTI and show that it has non-Abelian quasiparticles. The physical symmetries are realized in this surface topological ordered state in a manner forbidden in a strictly two dimensional insulator with the same topological order. 

The prime impetus for our study comes from recent progress in describing bosonic SPT phases in three dimensions. For bosons, interactions are essential to obtain an insulator. Consequently the study of boson SPTs is necessarily non-perturbative in the interaction strength. For such bosonic SPT phases, it was shown that the surface can be both gapped and symmetry preserving\cite{AshvinSenthilSPT} if it possesses intrinsic two-dimensional surface topological order (STO).   This STO however realizes symmetry in a manner prohibited in strictly two dimensional systems. The STO provides a particularly simple  non-perturbative insight into the bulk SPT phase.  Indeed targeting such an STO is a useful conceptual tool for constructing SPT phases\cite{ChongSenthil,burnellbc}, and can provide very general constraints on lower-dimensional phases\cite{AshvinSenthilSPT,ChongSenthil}.  In light of the simplicity and power of the STO as a surface termination of strongly interacting bosons SPTs it is natural to construct the STO appropriate for the fermionic topological insulator. We note that for fermionic topological superconductors in 3D protected by TRS very recently possible STO phases were constructed\cite{3DSC}.

Our strategy is to start from the TR-symmetric non-Abelian surface superconductor\cite{FuKane}, and to restore $U(1)_C$ without destroying the superconducting gap by proliferating vortices in the superconducting phase.  The minimal $\frac{hc}{2e}$ superconducting vortices cannot be directly condensed due to their non-Abelian statistics arising from unpaired core Majorana modes.  It turns out that, despite being Abelian, the doubled $\frac{hc}{e}$ vortex is a semion and can also not be condensed while preserving TRS.  Identifying an appropriate  vortex field that can be condensed to disorder the superconductor without breaking TRS requires some care. We find that there are $4$-fold  ($\frac{2hc}{e}$) vortex fields that can be condensed without breaking any symmetries as a minimal route to producing the STO starting from the surface superconductor. 

The resulting phase has identical topological order and charge assignments as the 2D Moore-Read quantum Hall state\cite{MooreRead} accompanied by an extra neutral semion.  However, in strictly two-dimensions this topological phase cannot be realized in a TR symmetric manner.  We will show that the fact that the eTI can realize this TO while preserving TRS provides a complete, non-perturbative definition of the bulk eTI.

\section{Vortex Condensation in a Conventional Superconductor}
As a warm-up for the more-complicated non-Abelian case, we begin by reviewing how insulating states can be produced by quantum disordering a conventional 2D s-wave superconductor through vortex proliferation.

A superconducting state has a charge $2e$ order parameter $\Delta = |\Delta| e^{2i\phi_s}$ that breaks $U(1)$ charge conservation symmetry. Starting from a conventional s-wave superconductor, one can restore $U(1)_C$ symmetry by proliferating vortices in the phase of the order-parameter, $\phi_s$.  Since the pairing amplitude $|\Delta|$ remains finite (except inside the vortex cores), the resulting state is clearly gapped.  Different gapped phases can be obtained by proliferating different types of vortices.  For example, proliferating $\pi$-vortices in $\phi_s$ (i.e. superconducting $\frac{hc}{2e}$ vortices) produces a simple band-insulator\cite{SCtoBandInsulator}, whereas proliferating $2\pi$ vortices produces a gapped phase with $\mathbb{Z}_2$ topological order\cite{SCtoBandInsulator,SCtoZ2}.  These constructions are well known\cite{SCtoBandInsulator,SCtoZ2}, but are useful to review in order to fix notation and to set the stage for the more complicated non-Abelian superconductors that are the subject of this paper.

All three phases are conveniently described by a parton construction in which the electron annihilation operator with spin-$\sigma$, $c_\sigma$, is rewritten as $c_\sigma = bf_\sigma$ with $b$ a spinless charge-1 boson (chargon) , and $f_\sigma$ is a neutral spinful fermion (spinon).  This parton description (often referred to as ``slave-boson"), has a $U(1)$ redundancy associated with changing the phase of $b$ and $f$ in opposite ways.  Consequently, any field theory description will contain an emergent, compact $U(1)$ gauge-field, whose vector potential we will denote by $a^\mu$.

\subsection{Superconductor}
In the parton description, the s-wave superconductor phase is described by condensing the charged boson, $\<b\>\neq 0$, and introducing an s-wave pairing amplitude for $f$: $\<f_\up f_\down\>\neq 0$.  In this phase, the emergent gauge field is gapped by the Higgs mechanism (or, equivalently confined) due to the charge-1 boson condensate.  

The gapped, unpaired $f$-quasiparticles are neutral fermion excitations.  These are ordinary Bogoliubov quasi-particles of the superconductor, that arise from electron states whose charge is screened completely (at long lengthscales) by the pair-condensate.

In addition, there are also $\pi$ vortices of the $f$-pair condensate phase.  Since $f$ carries internal gauge charge, these vortices carry $\pi$ ``magnetic"-flux of $a$.  The bosons, having internal gauge charge, are also affected by this $\pi$ flux of $a$.  Writing $b = \sqrt{\rho_b}e^{i\phi_b}$, we see that $\phi_b$ must wind by $\pi$ in the vicinity of this vortex in order to avoid an extensive energy penalty.  Since $b$ carries the physical electromagnetic charge, this means that a $\pi$-vortex in the f-pair-condensate is necessarily accompanied by a physical supercurrent flow in the $b$-condensate; this object is simply the familiar $\frac{hc}{2e}$ superconducting vortex.

\subsection{Band-Insulator}
In an s-wave SC, $\frac{hc}{2e}$ vortices carry only gapped quasi-particle states in their core.  Moreover, the pairing amplitude, $\Delta$, is non-vanishing outside of vortex cores.  Consequently, a state with an arbitrary density of non-overlapping vortices has no gapless excitations.  Therefore, one can consider starting with a superconductor and creating a quantum superposition of states with various numbers and placements of (well-separated) $\frac{hc}{2e}$ vortices.  This state will clearly be gapped.

Moreover, since the spinon excitations of the superconductor see the $\frac{hc}{2e}$ superconducting vortices as $\pi$-gauge-magnetic-flux, the spinon and vortex have mutual semionic statistics.  This immediately implies that the spinons will be confined in the vortex-proliferated state. The bosonic particles, $b$, also see the vortices as $\pi$-fluxes.  Therefore, the physical electron $c=bf$ has trivial mutual statistics with the vortex, remains gapped but deconfined.  Therefore, there is no spin-charge separation and the resulting state describes a conventional electron phase.

This phase can be thought of as a Bose-Mott insulator of Cooper pairs.
If the electron density is commensurate such that there are an even number of electrons per unit cell, then the Cooper pairs have integer filling and can form a Mott insulating state without further breaking any spatial symmetry.  Commensurate Cooper-pair filling is a necessary requirement for forming a band-insulator, and furthermore, the $\frac{hc}{2e}$-vortex-proliferated state has all the properties of an ordinary electronic band-insulator.  

Therefore, we see that $\frac{hc}{2e}$ vortex proliferation in a superconductor produces a conventional band-insulator.  This description of a band-insulator is clearly more complicated than the usual non-interacting band-structure description.  However, this construction provides a complementary ``dual" perspective capable of capturing correlated band-insulators, and can be a useful conceptual starting point for constructing more complicated strongly interacting phases.

\subsection{$\mathbb{Z}_2$ Topological Order}
Instead of proliferating $\frac{hc}{2e}$-vortices in the superconductor, one could alternatively proliferate doubled ($\frac{hc}{e}$) vortices.  If the electrons are at commensurate filling with the lattice, this proliferation destroys the boson superfluidity ($\<b\>=0$) without further breaking any other symmetry.  Single $b$-particle excitations are now gapped and the resulting phase is a charge insulator.  In this phase $a$ is not confined; rather, the emergent $U(1)$ gauge invariance is broken down to a local $\mathbb{Z}_2$ gauge invariance by the $f$-pair-condensate.  Moreover, since the spinons-$f$ develop a trivial (multiple of $2\pi$) Berry phase upon encircling an $\frac{hc}{e}$ defect, they remain deconfined.

The excitations of the theory are then $b$, $f$, and objects with $\pi$-flux of $a$ (visons).  The visons are their own antiparticles (since two visons make up the condensed $\frac{hc}{e}$ vortex), having mutual $\pi$-statistics with $b$ and $f$, and the resulting state is fractionalized with $\mathbb{Z}_2$ topological order.

It is worthwhile to pause to reflect on the strategy underlying the vortex condensation route to describing insulators proximate to superconducting phases in two space dimensions. In general a useful effective field theory description of such a system is formulated in terms of degrees of freedom natural in the superconductor - namely the $\frac{hc}{2e}$ vortices and the neutralized Bogoliubov quasiparticles (the $f$ field). The $\frac{hc}{2e}$ vortex field is a mutual semion with the $f$ particle and furthermore is coupled to a non-compact $U(1)$ gauge field. The vortex field of this  dual Landau-Ginzburg theory is, in the examples reviewed above, bosonic.  Vortex fields with strength $\frac{nhc}{2e}$ can therefore be formally condensed to produce various kinds of insulating states.

Having reviewed the simpler s-wave SC case, we now turn to the problem of producing a topologically ordered phase from the eTI surface-SC.

\section{Vortices in the eTI Surface Superconductor}
Starting from the superconducting surface of the eTI, we know that there should be some obstruction to proliferating superconducting vortices to form an ordinary band-insulator, and indeed the $\frac{hc}{2e}$-vortices in the superconducting TI surface-state are non-Abelian objects that cannot be directly condensed\cite{FuKane}.  Since $\frac{hc}{e}$ vortices do not have an unpaired Majorana core state, they are Abelian, and one is tempted to follow the above construction to obtain a $\mathbb{Z}_2$ topologically ordered state by proliferating $\frac{hc}{e}$ vortices.  

However, this naive approach fails to produce a symmetric STO state. It turns out that in the eTI surface SC, $\frac{hc}{e}$ vortices have semionic self-statistics\cite{Endnote:VortexStatistics}, and cannot be condensed without breaking TRS.  The $\frac{3hc}{2e}$ vortices again have unpaired Majorana cores, and are non-Abelian.  We show however that there are $\frac{2hc}{e}$ vortices that are bosonic. Therefore, the minimal route to restoring $U(1)_C$ is to condense such bosonic $\frac{2hc}{e}$ vortices. 

We now establish the Abelian statistics of $\frac{hc}{e}$ and $\frac{2hc}{e}$ vortices in the surface-superconductor, by arguing based on the $\Theta$-term electromagnetic response of the bulk.

\subsection{Bulk Argument for statistics of Abelian vortices}
\label{semi}
A useful conceptual device for what follows is to modify the problem by coupling the electrons to a weakly fluctuating dynamical compact $U(1)$ gauge field. It is well known that the 
topological insulating bulk leads to a $\Theta$-term, with $\Theta = \pi$,  in the effective action (apart from the usual Maxwell term) for this gauge field obtained by integrating out the electrons. Also well-known is the effect of this $\Theta$ term: a unit strength 
magnetic monopole of this $U(1)$ gauge field acquires electric charge $\frac{1}{2}$ (the Witten effect\cite{witten}). Now imagine tunneling such a monopole from the vacuum into the bulk of the (gauged) topological insulator. Such a tunneling process will leave behind at the surface a $\frac{hc}{e}$ vortex. This implies that the $\frac{hc}{e}$ vortex field in the vortex Landau-Ginzburg theory formally also has electric charge $\frac{1}{2}$.  As a composite made of charge-$1/2$ and $2\pi$ flux it is natural to expect that this vortex will have semionic statistics.  

\begin{figure}[ttt]
\includegraphics[width=2.2in]{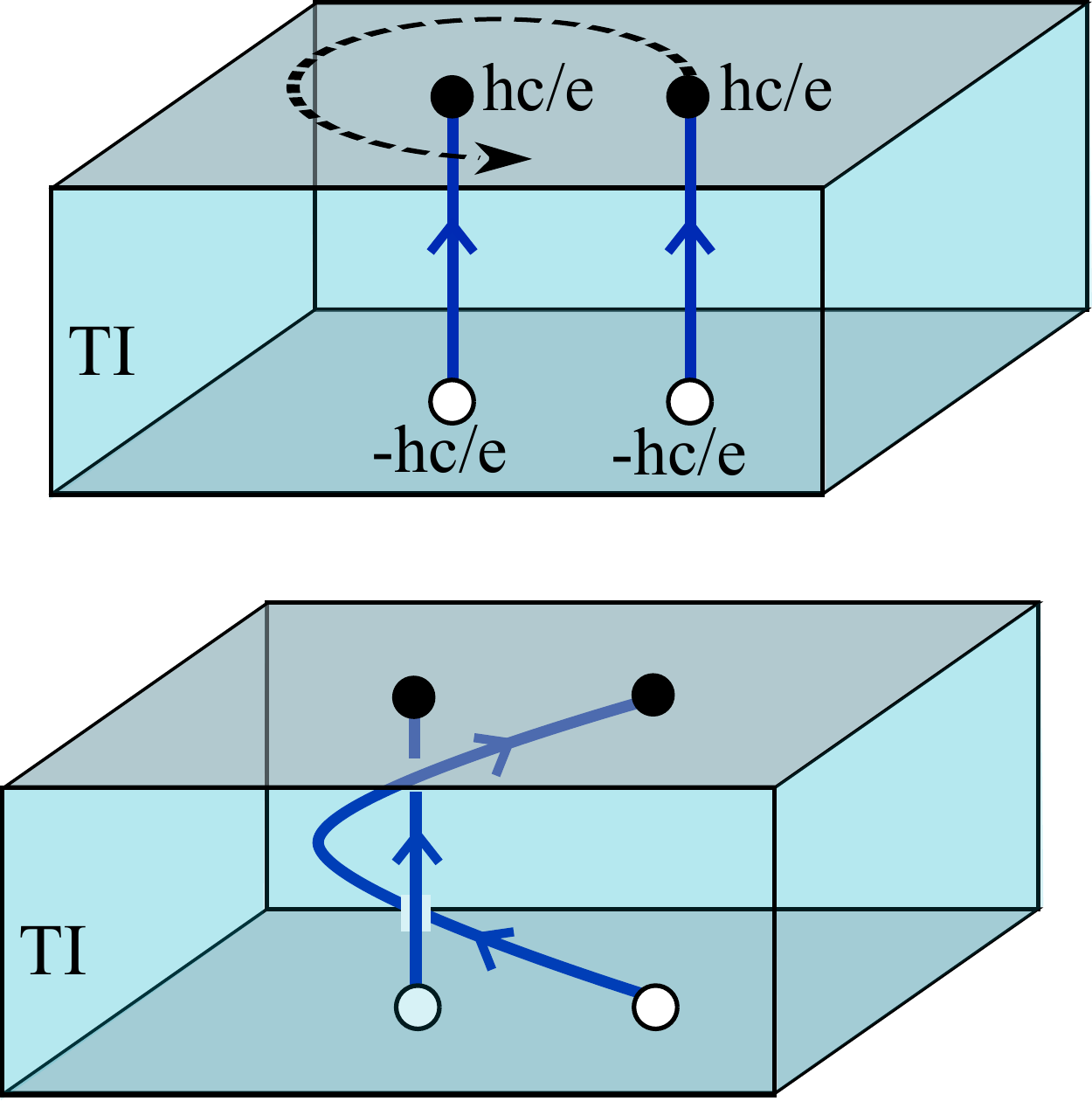}
\caption{Exchanging two $\frac{hc}{e}$ vortices at the superconducting surface of a TI slab (top panel) leads to a linking of their magnetic field lines, which gives a phase of $-1$, demonstrating that $\frac{hc}{e}$ vortices are semionic.}
\label{fig:VortexExchange}
\end{figure}

To demonstrate the semionic statistics of $\frac{hc}{e}$ vortices, consider a slab of bulk eTI with a top and bottom interface with a trivially insulating vacuum.    Then create a pair of $\frac{hc}{e}$ vortices on the top surface and a pair of $-\frac{hc}{e}$ vortices on the bottom surface.  Since the gauge field $A^\mu$ is free, except at the superconducting surface, closed magnetic flux lines carrying $\frac{hc}{e}$ flux are condensed in the bulk and in the vacuum.  Since the surface is superconducting, a magnetic flux tube can only penetrate the surface at a vortex.  For the vortex configuration of Fig.~\ref{fig:VortexExchange}, there are only two magnetic flux lines that leave the TI bulk.  Let us consider just one representative flux line configuration, as shown in Fig.~\ref{fig:VortexExchange}.  Next consider dragging one of the $\frac{hc}{e}$ vortices on the top surface all the way around the other, as shown in Fig.~\ref{fig:VortexExchange} without moving the $-\frac{hc}{e}$ vortices on bottom surface.  The new magnetic flux configuration differs from the initial one by a single linking of the magnetic flux lines that thread the vortices.

Due to the bulk topological $\Theta$-term for $A$: 
\begin{align} \mathcal{L}_\Theta = i\frac{\e^{\mu\nu\lambda\rho}}{8\pi}\partial_\mu A_\nu \partial_\lambda A_\rho \end{align}
this linking produces a phase of $-1$ relative to unlinked configurations.  This phase can be computed directly from $\mathcal{L}_\Theta$ by considering any convenient choice of $A$ with a linked vortex line. Alternatively, one can imagine creating a linked field line configuration in the bulk by starting with an infinite flux line, creating a monopole anti-monopole pair and dragging the monopole around the flux line before annihilating it with the anti-monopole.  Since monopoles in the TI bulk have charge $\frac{e}{2}$, dragging one around a $2\pi$-flux line contributes phase $e^{2\pi i \cdot \frac{1}{2}}=-1$.  

We have illustrated this $-1$ phase for a particular magnetic field line configuration.  More generally, the ground-state, $|\Psi_\text{EM}\>$, of the bulk gauge field, $A^\mu$, is a quantum-superposition of various configurations, $\mathcal{C}$, of magnetic flux lines: 
\begin{align}
|\Psi_\text{EM}\> = \sum_\mathcal{C} (-1)^{L_\mathcal{C}}\Psi_0(\mathcal{C})|\mathcal{C}\>
\end{align}
weighted by phase $(-1)^{L_\mathcal{C}}$, where $L_\mathcal{C}$ is the number of linked loops in the configuration $\mathcal{C}$, and by amplitude, $\Psi_0(\mathcal{C})$, that is determined by the non-topological dynamical terms for the gauge-field.

This follows directly from computing the wave-function for a given configuration, $\mathcal{A(\v{r})}$,  from the (imaginary time) path integral: 
\begin{align} \Psi\[\mathcal{A}\]&= \<\mathcal{A}|\Psi\> 
\nonumber\\
&= \int D[A]\big{|}_{A_\mu(\v{r},t=0)=\mathcal{A}(\v{r})}e^{-\int_{-\infty}^0d\tau \int d^3r\mathcal{L}_\Theta[A]} 
\nonumber\\
&\sim e^{i\int d^3r\frac{\e^{\mu\nu\lambda}}{8\pi}\mathcal{A}_\mu\partial_\nu \mathcal{A}_\lambda} =(-1)^{L_\mathcal{C[A]}} \end{align}
we see that the resulting wave-function contains a Chern-Simons (CS) term  which just counts the linking number of flux lines of $\mathcal{A}$.  

For any configuration of closed bulk field-lines,  $\mathcal{C}$, the two-fold exchange of $\frac{hc}{e}$ introduces a single extra linking number.  Therefore the two-fold exchange of $\frac{hc}{e}$ vortices produces phase $(-1)$, indicating that a single exchange produces phase $\pm i$; the $\frac{hc}{e}$ vortices are semionic.  Let us denote the quantum field that creates a $\frac{nhc}{e}$ vortex with electric charge $q$  by $\Phi_{n,  q}$. With this notation $\Phi_{1, \frac{1}{2}}$ is a semionic $\frac{hc}{e}$ vortex with charge $\frac{1}{2}$.  The field $f\Phi_{1, \frac{1}{2}}$ produces a neutral fermion bound to this vortex and hence creates an antisemionic $\frac{hc}{e}$ vortex with charge $\frac{1}{2}$. These two $\frac{hc}{e}$ vortices will play an important role below. 

Let  us now consider strength-$4$ ($\frac{2hc}{e}$) vortices. A similar argument as above shows that $\frac{2hc}{e}$ vortices are either bosonic or fermionic (fermionic and bosonic $\frac{2hc}{e}$ vortices can be interchanged by binding a neutral $f$ quasi-particle).  Note that if we combine two charge-$1/2$ semionic $\frac{hc}{e}$ vortices, we end up with a charge-$1$ bosonic  $\frac{2hc}{e}$ vortex. {\em i.e} $(\Phi_{1, \frac{1}{2}})^2 = \Phi_{2,1}$.  An electrically  {\em neutral} $\frac{2hc}{e}$ vortex may be obtained by  considering the combination $c \Phi_{2,1}$, {\em i.e} by removing an electron from the charge-$1$ $\frac{2hc}{e}$ vortex. Clearly this is a fermion.

These strength-$4$ vortices at the surface correspond in the bulk to strength-$2$ monopoles. At $\Theta = \pi$, such monopoles always carry integer electric charge.  
 We will denote bulk dyons with magnetic charge $n$ and electric charge $q$ by $(n,q)$. These correspond to surface vortices created by $\Phi_{n,q}$. It is readily seen that the bulk $(2,1)$ dyon (at $\Theta = \pi$) is a boson while the electrically neutral strength $2$ monopole  (the $(2,0)$ particle) is clearly a fermion 
(the polarization charge induced by the $\Theta$ term does not contribute to the statistics, as explained in Ref.\onlinecite{dyonstatistics}). This is in complete accord with our discussion of surface vortices above.  Arguments using bulk monopole properties to constrain surface physics were also recently used for boson topological insulators in Ref.~\onlinecite{statwitten}. 

To disorder the surface superconductor we need to identify bosonic vortices which we can then condense. Though the $\frac{2hc}{e}$ vortex with electric charge-$1$ seems like a candidate it is problematic. To preserve time reversal we should clearly also condense (with equal amplitude) the $-\frac{2hc}{e}$ vortex with electric charge $1$. But then the resulting state also has a condensate of ordinary Cooper pairs so that it is still a superconductor (albeit an exotic one). The neutral $\frac{2hc}{e}$ vortex described above is a fermion and hence cannot condense. 
Fortunately we also have a different neutral fermion in our theory - the spinon (the $f$ particle). By binding $f$ to the fermionic $\frac{2hc}{e}$ vortex we obtain an electrically neutral bosonic $\frac{2hc}{e}$ vortex.  Equivalently this bosonic neutral $\frac{2hc}{e}$ vortex may be viewed as being obtained from the charge-$1$ bosonic $\frac{2hc}{e}$ vortex by binding to $b$ ({\em i.e} by removing a chargon). This neutralizes the charge but keeps the statistics as bosonic. We are then free to condense this vortex to destroy the superconducting order.

We emphasize that the bosonic neutral $\frac{2hc}{e}$ vortex is not simply a $4\pi$ vortex of the chargon $b$ but requires also binding to the spinon $f$.  An $8\pi$ ($\frac{4hc}{e}$) vortex of $b$, $\Phi_{4,0}$,is an electrically neutral boson. The corresponding bulk monopole is a $(4,0)$ particle which is also a boson. Condensation of 
the bosonic vortex $f\Phi_{2,0}$ automatically implies condensation of  $\Phi_{4,0}$ as the spinon $f$ is paired. 

\subsection{Topological spins of non-Abelian vortices}
\label{nav}

We now consider non-Abelian vortices, and it is sufficient for our purpose to consider $\pm hc/2e$ vortices, with Majorana core states. 
Naively, the argument given in Sec.\ref{semi} for $hc/e$ vortices implies that the topological spin (see Sec.\ref{ts} for its definition) of $\pm hc/2e$ vortices would be $e^{i\pi/8}$. This can be seen by writing the 
bulk $\Theta$-term as a boundary Chern-Simons term at level-$1/2$, which would contribute to the topological spin of $\pm hc/2e$ vortices by $e^{i\pi/8}$. 
However, the Majorana zero-modes trapped in the vortices contributes another $e^{-i\pi/8}$ to the topological spins\cite{KitaevHoneycomb}, 
hence the total topological spins are
\begin{equation}
\label{navts}
 \theta_{hc/2e}=\theta_{-hc/2e}=1.
\end{equation}
 
The above argument can be made more precise by viewing the surface superconductor as a paired single Dirac cone. One can then add two gapped Dirac cones with opposite masses to the surface without breaking time-reversal symmetry. 
One can then group one of the massive Dirac cones with the original surface superconductor and rewrite the combination as a $p-ip$ superconductor, and the other massive Dirac cone with the opposite mass gives a half-quantum hall state. 
The former contributes $e^{-i\pi/8}$ to the topological spin of $\pm hc/2e$ vortices, and the latter gives $e^{i\pi/8}$, hence we have $\theta_{\pm hc/2e}=1$.

\section{Surface Topological Order}
We are now in a good position to construct a symmetry preserving STO phase from the SC phase.  In the parton construction $c_\sigma = bf_\sigma$, we can describe the SC topological insulating surface state by condensing b, $\<b\>\neq 0$, and placing $f$ in the eTI band-structure with a superconducting surface.  From the previous section, we saw that the minimal route to restoring the $U(1)_C$ symmetry is to proliferate the electrically neutral bosonic $\frac{2hc}{e}$ vortices.  

What topologically distinct classes of particles remain after their proliferation?  Since $b$ and $f$ have trivial mutual statistics with the $\frac{2hc}{e}$ vortices, they will clearly survive as gapped quasi-particles with unaltered charge and statistics.  Quite generally the condensation of such $\frac{2hc}{e}$ vortices will produce an insulator with gapped bosonic excitations with fractional charge $1/2$. We will call this particle $\beta$. Clearly two $\beta$ particles make a chargon: $b = \beta^2$.

Vortices in the superconductor become dressed by the $\frac{2hc}{e}$ condensate.  We will see later that they survive as topological quasiparticles but with  sharp non-zero electric charge (unlike in the example reviewed above of 2D $\mathbb{Z}_2$ topologically ordered states produced by disordering a proximate superconductor, where the visons are charge neutral).  For now, we put aside the charge assignment for these topological particles and focus just on identifying the different particle types. 

Going from the superconductor to the STO phase, the non-Abelian $\frac{hc}{2e}$ vortex, $v$, becomes a new object, $\tau_v$, which is a quantum superposition of odd-strength vortices in the superconductor whose vorticity differs by a multiple of $\frac{2hc}{e}$.  Similarly, the $-\frac{hc}{2e}$ anti-vortex, $\bar{v}$, becomes a different object, $\tau_{\bar{v}}$, which is made up of a quantum superposition of $\frac{(4n-1)hc}{2e}$ vortices of the superconductor (with $n\in\mathbb{Z}$).

In the SC, an $\frac{hc}{2e}$ vortex, $v$, carries a Majorana zero mode in its core\cite{FuKane}, and a pair of $v$'s shares a single complex fermion level that can be either occupied or unoccupied.  Consequently, there are two possible outcomes from fusing two $v$'s, $v^2_\pm$, both of which have net vorticity $\frac{hc}{e}$ and which differ from each other by adding a neutral Bogoliubov fermion, $f$.  Upon moving into the STO phase by condensing 4-fold vortices, $v^2_\pm$ will turn into distinct objects, $\tau^2_\pm$, which differ by a fermion: $\tau^2_+=\tau^2_-\times f$.

Similarly, in the superconductor, a pair of $\bar{v}$'s can fuse to two different $-\frac{hc}{e}$ vortex objects that differ by a fermion, $f$.  Upon condensing $\frac{2hc}{e}$ vortices however, the $\pm\frac{hc}{e}$ vortices become mixed, and fusing two $\tau_{\bar{v}}$ particles should have the same outcome as fusing two $\tau_v$ particles: $\tau_{\bar{v}}\times\tau_{\bar{v}} = \tau_v\times\tau_v= \tau^2_+ +\tau^2_-$. 

Lastly, in the superconductor, the vortex and anti-vortex pair also share a non-local fermion level due to their Majorana cores.  Fusing a $v$ and $\bar{v}$, then produces either the superconducting ground state, $\mathbb{I}$, or the ground-state plus an extra Bogoliubov particle: $v\times\bar{v} = \mathbb{I}+f$.  Consequently, in the STO phase, we must have two possible fusion outcomes for $\tau_v\times\tau_{\bar{v}}$, which differ by an $f$.  Naively, one might be tempted to have $\tau_v$ and $\tau_{\bar{v}}$ fuse $1+f$ as in the superconductor.  However, more generally we may also have: $\tau_v\times\tau_{\bar{v}} = X\times(1+f)$ where $X$ is some to-be determined particle that is condensed in the SC.  This is consistent with the fusion rules of the surface SC if $X$ is condensed in the SC phase.  This requires $X$ to be a boson.  Below we will show that $X$ is just the fractional chargon: $\beta$.

Finally, we note that $\tau^2_+\times\tau^2_+=\beta^2$, and that $\tau^2_+\times\tau^2_-=\beta^2\times f=c$, the physical electron.

A summary of the particle content and fusion rules produced by this line of reasoning is summarized in Tables ~\ref{Tbl:TopoProperties} and \ref{Tbl:FusionRules} respectively.

\renewcommand{\arraystretch}{1.4}
\setlength{\tabcolsep}{7pt}
\begin{table*}[tb]
\begin{tabular}{|>{\centering\arraybackslash}m{1.5in}|>{\centering\arraybackslash}m{.5in}cccccccc|}
\hline
Topological Superselection Sector (``Particle Type"): & $\mathbb{I}$ & $\beta$ & $f$ & $\tv$ & $\tvb$ & $\tau^2_+$ & $\tau^2_-$ &$\tau^3_v$&$\tau^3_{\bar{v}}$
\\ \hline
Conjugate Sector (anti-particle): & $\mathbb{I}$ & $\beta^3\equiv\beta^{-1}$&$f$&$\beta^{-1}\tvb$&$\beta^{-1}\tv$& $\beta^{-2}\tau^2_+$&$\beta^{-2}\tau^2_-$ &$\beta^{-2}\tv$&$\beta^{-2}\tvb$
\\ \hline
Quantum Dimension (d): & 1&1&1&$\sqrt{2}$&$\sqrt{2}$&1&1&$\sqrt{2}$&$\sqrt{2}$
\\ \hline
Topological Spin ($\theta$):& 1&1&-1&$e^{i\pi/4}$&$e^{-i\pi/4}$&$e^{i\pi/2}$&$e^{-i\pi/2}$&$-e^{i\pi/4}$&$-e^{-i\pi/4}$
\\ \hline 
Charge ($q_e$):& $2ne$\newline ($n\in\mathbb{Z}$) & $\frac{e}{2}$&0&$\frac{e}{4}$&$\frac{e}{4}$&$\frac{e}{2}$&$\frac{e}{2}$&$\frac{3e}{2}$&$\frac{3e}{2}$
\\ \hline 
Time-Reverse Partner:& $\mathbb{I}$ & $\beta$ & $f$ & $\tvb$ & $\tv$ & $\tau^2_-$ & $\tau^2_+$& $\tau^3_{\bar{v}}$& $\tau^3_v$
\\ \hline
$\T^2$ value (if meaningful):&1&1&-1&&&&&&
\\ \hline
\end{tabular}
\caption{Summary of the topological content of the surface-topological order phase and the implementation of charge-conservation and TR symmetries. Topological superselection sectors are topological equivalence classes of particle types.  The anti-particle of a particle in sector $a$ resides $a$'s conjugate sector.  A particle has the same quantum-dimension and $\T^2$ value as its anti-particle, but opposite electrical charge and conjugate topological spin. Other distinct topological particles such as $\beta^2$, $\beta\tv$, etc... can be obtained by combining the above listed objects.  The properties of these composites and anti-particles follows straightforwardly from the information listed above.  Superselection sectors have the same quantum dimension, opposite charge, and same topological spin compared to their conjugate sectors (anti-particles).  Empty entries in the $\T^2$ row indicate that there is no gauge invariant meaning to the value of $\T^2$ for that type of particle.  In addition, there is the physical electron, $c$, which has $d=1$, $\theta_c=-1$, $\T^2_c=-1$.  This could be regarded as part of the vacuum sector $\mathbb{I}$ since it has trivial mutual statistics with all other particles.  However, since fusing $c$ to another particle changes that particle's topological spin  factor of $-1$ it is convenient to distinguish $c$ from $\mathbb{I}$.}
\label{Tbl:TopoProperties}
\end{table*}

\begin{table}[tbh]
\begin{tabular}{|c|}
\hline
$\beta\times\beta = \beta^2$ \\
$\beta^2\times\beta = \beta^3 = \beta^{-1}$ \\
$\beta^3\times\beta = \mathbb{I}$ \\
$\beta^n\times a= \beta^n a$ \\
(for any sector $a\neq\beta$ and $n=1,2,3$)\\
\hline
$f\times f = 1$ \\
$f\times\tv = \tv$ \\
$f\times\tvb=\tvb$ \\
\hline
$\tv\times\tv=\tau^2_++\tau^2_-$ \\
$\tvb\times\tvb = \tau^2_++\tau^2_-$ \\
$\tv\times\tvb = \beta+\beta f$ \\
\hline
$\tau^2_\pm\times f = \tau^2_\mp$\\
$\tau^2_\pm\times\tau^2_\pm = \beta^2$ \\
$\tau^2_+\times\tau^2_-=\beta^2\times f = c$ \\
\hline
$\tau^3_v=\tau_v\times\tau_\pm^2$\\
$\tau^3_{\bar{v}}=\tau_{\bar{v}}\times\tau_\pm^2$\\
\hline
\end{tabular}
\caption{Fusion rules for the surface-topological order phase.}
\label{Tbl:FusionRules}
\end{table}

\subsection{Charge Assignments}
Having specified the topologically distinct particle classes and fusion rules for the STO phase, we now turn to their symmetry properties under $U(1)_C$.  The resulting charge assignments explained below are summarized in Table ~\ref{Tbl:TopoProperties}.

Since $b$ and $f$ are unaffected by the vortex condensation, $b$ still carries charge $e$ and that $f$ is charge-neutral.  What about the excitations that descend from superconducting vortices?  $\tau^2_\pm$ particles descend from $\Phi_{1,1/2}$ vortex fields of the superconductor, and hence can be created by dragging a magnetic monopole from the vacuum through the STO surface into the bulk. Since the monopole carries fractional electric charge: $\pm\frac{e}{2}$, its corresponding surface excitations must also have charge $\mp\frac{e}{2}$.  Moreover, since $\tau^2_+$ and $\tau^2_-$ differ by a neutral fermion, $f$ they must have the same charge.  For concreteness, and without loss of generality, we choose $\tau_\pm^2$ to have charge $+\frac{e}{2}$ and their anti-particles, $\tau^{-2}_\pm$, to have charge $-\frac{e}{2}$. It then immediately follows from the fusion rule: $\tau_v\times\tau_v = \tau^2_+ +\tau^2_-$ that $\tau_v$ has charge $\frac{e}{4}$.  

It is instructive to understand how these charge assignments come about directly from the surface without recourse to bulk monopoles. To obtain the STO from the SC, we are condensing $4\pi$ vortices of the chargon $b$ that are bound to the neutral fermion $f$.  The neutral fermion acquires a $\pi$ phase when it encircles the $\frac{hc}{2e}$ vortex in the superconductor.  Consequently, the $\frac{hc}{2e}$ vortex is a mutual semion with the condensed bosonic $\frac{2hc}{e}$ vortex.  As a result, the $\frac{hc}{2e}$ vortex can survive in the STO phase only by binding with some other particle to produce trivial mutual statistics with the condensed bosonic $\frac{2hc}{e}$ vortex.  The only possibility is for the $\frac{hc}{2e}$ vortex to bind a fractional charge, $\frac{e}{4}$, which also obtains $\pi$-phase upon encircling an $\frac{2hc}{e}$ vortex. Thus we conclude that the particles $\tau_v, \tau_{\bar{v}}$ in the STO phase are the remnants of the $\frac{hc}{2e}$ vortices of the SC phase which have been dressed by charge $e/4$. 

Since $\tau_v$ and $\tau_{\bar{v}}$ descend from $\pm\frac{hc}{2e}$ vortices in the superconductor, they are related by time-reversal and must have the same charge.  Above, we saw that the $v\times\bar{v}=1+f$ fusion rule for the surface-SC generalized to: $\tau_v\times\tau_{\bar{v}}= X\times(1+f)$ in the STO phase, with $X$ to-be-determined particle.  The above arguments show that $X$ must have charge $\frac{e}{2}$.  Since $X$ is a $\frac{1}{2}$-charge boson that must be condensed in the SC phase, the only possibility is: $X=\beta$.

\subsection{Topological Spins}
\label{ts}
The topological spin, $\theta_a$, of a particle in sector $a$ is defined as the phase factor accumulated when an $a$-particle is adiabatically rotated by $2\pi$ in the counter-clockwise (CCW) sense.  For Abelian particles, the topological spin coincides with the phase obtained through CCW exchange of a pair of $a$-particles.  

Clearly $\theta_b=1$ and $\theta_f=-1$. 
The argument in Sec.\ref{semi} established the semionic/anti-semionic statistics of $hc/e$ surface vortices (the semion and anti-semion differ by an $f$ fermion). In the topologically ordered phase the $hc/e$ vortex acquires an additional charge $q_{\tau^2_{\pm}}=1/2$. The charge-flux relation thus gives an additional $e^{iq\phi}= -1$ to its topological spin. 
This shifts a semion to an antisemion and vise versa. But since we have both semionic and anti-semionic vortices already, the shift is just a relabeling of the two different vortices. Hence we establish that $\tau^2_{\pm}$ have topological spin $\pm i$.

It was also established in Eq.\ref{navts} that the $\pm hc/2e$ vortices have trivial topological spins. In the topologically ordered phase, the $\pm hc/2e$ vortices acquire additional charge-$1/4$ and becomes $\{\tau_v,\tau_{\bar{v}}\}$. 
Hence an additional contribution of $e^{iq\phi}=e^{\pm i\pi/4}$ is introduced to the topological spin. Hence we have $\theta_{\tau_v}=e^{i\pi/4}$ and $\theta_{\tau_{\bar{v}}}=e^{-i\pi/4}$.

\subsection{Exchange Statistics}
In a system with non-Abelian particles that have multiple possible fusion outcomes, the phase obtained by the CCW exchange of two particles, $a$ and $b$, will depend on the fusion channel.  When $a$ and $b$ fuse to $c$, the phase factor obtained by adiabatic CCW exchange of $a$ and $b$ is denoted by $R^{ab}_c$ (for a pedagogical review see Ref.~\onlinecite{PreskillNotes}).  The $R$ matrices are related to the topological spin of the underlying particles\cite{PreskillNotes} by $(R^{ab}_c)^2 = \theta_c/\theta_a\theta_b$.  This identity just encodes the fact that dragging $b$ around $a$ is nearly the same as rotating the entire $a$-$b$ composite system CCW by $2\pi$, or equivalently to fusing to $c$ and rotating CCW by $2\pi$ giving: $\theta_c$.  However, rotating the entire system also rotates $a$ and $b$ individually, which is not part of the exchange process.  The factor of $\theta_a\theta_b$ in the denominator compensates for this unwanted rotation of $a$ and $b$.  The proper branch of the square-root can be identified by writing $\theta_{a,b,c}\equiv e^{i\phi_{a,b,c}}$, and choosing an exchange protocol such that the phase is accumulated monotonically over the course of time T: $R^{ab}_c = \underset{{t\rightarrow T^{-}}}{\lim}e^{i(\phi_c-\phi_a-\phi_b)t/2T}$.

For Abelian particles $a$ and $b$, there is a unique fusion channel, and the lower-index on $R$ is redundant.  Therefore, it is common to just specify the mutual statistics of $a$ and $b$ by: $\theta_{a,b} = (R^{ab}_{a\times b})^2$, which is the phase factor obtained by adiabatically dragging $b$ CCW around $a$.  Consequently, the braiding statistics for all particles follows straightforwardly from the previously obtained fusion rules and topological spins tabulated in Tables.~\ref{Tbl:TopoProperties} and \ref{Tbl:FusionRules} respectively.

For example, consider the mutual statistics of $\tau_v$ and $\tau_\pm^2$.  The composite $\tau_v\times\tau_\pm^2 = \tau_v^3$ has topological spin: $\theta_{\tv^3} = -e^{i\pi/4}$, indicating:
\begin{align} \theta_{\tv,\tau_\pm^2} = \frac{\theta_{\tv\times\tau_\pm^2}}{\theta_{\tv}\theta_{\tau_\pm^2}} = \frac{-e^{i\pi/4}}{e^{i\pi/4}e^{\pm i\pi/2}} = -e^{\mp i\pi/2}
\end{align}

\subsection{Time-Reversal Properties}
We have already identified appropriate charge assignments, which encode the transformation properties of various particles under the $U(1)_C$ symmetry.  In this section, we address how TR is implemented in the proposed STO phase.  The results of this section are summarized in Table.~\ref{Tbl:TopoProperties}.  

The first task for implementing TRS is to specify how topological equivalence classes of particles are exchanged under TR.  This is relatively straightforward since we have constructed the STO state from the well-understood TR-symmetric superconductor phase.  The $\tv$ descends from an $\frac{hc}{e}$ vortex in the superconductor, which becomes a $-\frac{hc}{e}$ vortex under TR; in turn the $-\frac{hc}{e}$ vortex becomes $\tvb$ in the STO phase.  Therefore under TR:
\begin{align} \tv\overset{\T}{\longleftrightarrow}\tvb \end{align}
Similarly, by going to the superconductor it is clear that $f$, and  $\beta^2\cong b$ are preserved under TR.  It is also clear that the $\beta$ sector is preserved under TR.

Under TR, counter-clock-wise and clock-wise exchange are interchanged, and hence topological classes of particles that are related by TR must have conjugate topological spin.  We see that this is true for all of the above TR transformation rules.

Since $\tau^2_\pm$ descend from both $\pm\frac{hc}{e}$ vortices, we cannot determine their TR properties directly from the superconductor. However, since $\tau^2_+$ and $\tau^2_-$ have conjugate topological spins, they must be exchanged by $\T$:
\begin{align} \tau^2_+\overset{\T}{\longleftrightarrow}\tau^2_- \end{align}

In addition to the action of $\T$ on topological superselection sectors, for sectors that are not interchanged by $\T$, it is meaningful to ask about their eigenvalues under the unitary operation of double-time-reversal, $\T^2$.  For particles that reside in TR-invariant superselection sectors, $\T^2=-1$ has definite physical interpretation as a TRS-protected Kramers degeneracy.  Our STO state arises naturally from the superconductor where $b$ has $\T^2=1$ and $f$ has $\T^2=-1$ respectively; hence $\beta^2$ and $f$ also have $\T^2=1$ and $\T^2=-1$ in the STO phase.  Similarly, $\beta^2$ has $\T^2=1$ since it is a fraction of $b$, and since $\beta$ can be condensed to obtain the SC from the STO phase.

However, for particles, like $\tau_\pm^2$, whose superselection sectors are changed by $\T$, the $\T^2$ eigenvalue does not imply a further degeneracy within that particle sector.  Furthermore, for such particles, it turns out that it is not even possible to assign a local gauge-invariant representation of $\T^2$. In the next two sections we further describe the issue of symmetry localization on gauge non-invariant particles. 

\subsubsection{Gauge (non)-invariance TR Properties for Fractionalized Particles}
Fractionalized particles (i.e. particles with non-trivial self- or mutual-statistics) cannot be individually created from the ground-state.  Rather, one can only create groups of excitations that fuse to $\mathbb{I}$.  For example, to isolate a fractionalized particle $X$, one can create a particle anti-particle pair, $X$ and $X^{-1}$, from the ground-state, and pull them far apart from one another.  The operator that implements this sequence consists of a string of electron operators connecting the final locations, $R_1$ and $R_2$, of $X$ and $X^{-1}$ respectively.  This string of operators can be divided into two local operators $\Psi^\dagger_X(R_1)$ and $\Psi_{X}(R_2)$, that create $X$ and $X^{-1}$ respectively, and a non-local gauge-string, $W_{1,2}=\prod_{\Gamma}e^{i q_X a_{ij}}$, where $i$ and $j$ label sites on the lattice where $\Psi_X$ is defined, $\Gamma$ is directed path of links $\<ij\>$ connecting sites $R_1$ to $R_2$, $q_X$ is the internal gauge-charge of the particle $X$, and $a_{ij}$ is a discrete-valued emergent gauge field.  This division into particles and strings is inherently arbitrary, which is reflected by the local gauge invariance under the transformations $\Psi_{X,i}\rightarrow e^{2\pi in_i q_x}\Psi_{X,i}$, and $a_{ij}\rightarrow a_{ij}-\(n_i-n_j\)$ (with $n_{i,j}\in\mathbb{Z}$).

Due to the non-local gauge structure there is not always a well-defined gauge invariant way to assign symmetry-transformation properties locally to the particle creation operators $\Psi^\dagger_X$.  Rather, one must generically keep track of the transformation property of both the particles, and their gauge-strings, $W$.  However, in special cases it is possible to associate a well-defined action of a symmetry locally on $\Psi^\dagger_X$ even for gauge non-invariant objects.  For simplicity, in what follows, we will not distinguish between the label $X$ for a topological class of particles and the corresponding (gauge-non-invariant) annihilation operator $\Psi_X$.

Since $f\times f=\mathbb{I}$, the phase of $f$ has a sign ambivalence, indicating that $f$'s have $\frac{1}{2}$-gauge charge (i.e. change sign under $e^{2\pi i q_f}=-1$) and are connected pairwise by (unobservable) unoriented $\mathbb{Z}_2$ gauge strings.  Similarly $\beta^2\times f$ is a physical electron $c$, and so $\beta^2$ also has a $\mathbb{Z}_2$ gauge charge.  It then follows from $\tau_\pm^2\times\tau_\pm^2=\beta^2$, that $\tau_\pm^2$ has internal $\frac{1}{4}$-gauge charge, and that oriented $\mathbb{Z}_4$ gauge strings emanate from $\tau^2_\pm$ particles.  We know that $\tau^2_\pm$ have opposite internal gauge charge, since $\tau^2_+\times\tau^2_-=c$, and $c$ is a physical (gauge-invariant) local electron.  Therefore, we can choose the orientation convention that $\mathbb{Z}_4$ lines emanate from $\tau_+^2$ and terminate on $\tau_-^2$ particles. 

\subsubsection{$\T^2$ Properties For Sectors that are Exchanged by $\T$}
With this gauge-string picture in mind, we now turn to the task of determining to what extent $\T^2$ is defined on particles whose topological classes are interchanged by $\T$.  To see why it is important to consider the effects of $\T$ on the gauge string, consider a $\tau_+^2$-$\tau_-^2$ pair.  Suppose that we represent $\T$ locally on the particle operators as: $\T^{-1}\tau^2_+\T = e^{i\alpha}\tau^2_-$ and $\T^{-1}\tau^2_-\T = e^{i\beta}\tau^2_+$ where $\alpha$ and $\beta$ are unknown phases. Then one has: $\T^{-2}\tau^2_\pm\T^2=e^{\pm i(\beta-\alpha)}$, and naively it appears that $\T^{-2}\tau^2_+\tau^2_-T^2 = |e^{i(\beta-\alpha)}|^2\tau^2_+\tau^2_-$.  However, this cannot be the whole story, since $\tau^2_\pm$ fuse to the physical electron, $c$, which is a Kramers doublet with $\T^2=-1$.

This puzzle is resolved by noting that $\tau_+^2\overset{\T}{\leftrightarrow} \tau_-^2$, implies that $\T$ reverses the direction of the gauge string connecting a given $\tau^2_+$-$\tau^2_-$ pair.  Then acting twice with $\T^2$ doubly flips the orientation of the connecting gauge-string.  A two-fold re-orientation of the gauge-string can also be accomplished by dragging $\tau^2_+$ around $\tau^2_-$.  Due to their semionic mutual statistics, this observation dictates that the gauge string contributes an additional factor of $-1$ to the overall $\T^2$.  Therefore, the action of $\T^2$ cannot be consistently implemented in a purely local fashion for the gauge-non-invariant particles $\tau^2_\pm$, which interchange under $\T$.

Note that a nearly identical argument can be applied to monopoles in the bulk of the electron TI to formally establish the intimate connection between the $\theta=\pi$ electromagnetic response of the TI and the Kramers degeneracy of the electron\cite{fSPTClassification} (see also Ref.~\onlinecite{Metlitski-eTI}).  This is indeed appropriate, since the $\tau^2_\pm$ particles are the surface-avatars of these bulk dyons.

The issue of non-locality is even more pronounced for the non-Abelian excitations $\tv$ and $\tvb$, since a collection of these particles share a degenerate Hilbert space of non-local fermion modes, and the action of $\T^2$ depends on the total fermion parity of this non-local Hilbert space, which is a global property of the system.

\section{2D TR Breaking Analog}
For bosonic SPT bulk phases, the topological properties of the STO phase can always be realized by a strictly 2D system that does not preserve the underlying symmetries of the 3D SPT.  In this section, we provide an analogous construction for the electron TI.  Specifically, we show that the STO phase has the same topological order as the Moore-Read QH phase\cite{MooreRead} supplemented by an extra neutral semion. We begin by reviewing the Moore-Read and related 2D phases in the language of the parton construction $c_\sigma=bf_\sigma$ used above.

\subsection{$p+ip$ Superconductor and Kitaev Spin-Liquid}
We begin with the $p+ip$ superconductor, and its topologically ordered analog, which are in some sense the simplest ``roots" of the non-Abelian Ising topological order for the STO phase.  A TR-breaking superconductor with $p+ip$ pairing symmetry, and the TR-broken B-phase of Kitaev's Honeycomb Model (henceforth denoted Kitaev Spin-Liquid, or KSL) are closely related states with non-Abelian Ising anyon excitations.  The latter is obtained from the former by condensing $\frac{hc}{e}$-vortices.  In the language of the parton construction, this is equivalent to placing $b$ in a Mott insulator, and $f$ into a $p+ip$ superconductor.  The resulting phase contains topological particle classes: $\mathbb{I}$ (vacuum), $b$, $f$, and a non-Abelian vison, $\sigma$ that descends from the $\pm\pi$-vortices of the $p+ip$ superconductor.

In the resulting KSL phase, $b$ has charge $e$, and all other particles are neutral.  The edge of this phase contains a single chiral Majorana fermion that contributes $\sigma_{H}=0$ and $\kappa_H=\frac{1}{2}$.  The fusion rules are:
\begin{align}
b\times f &= c \\ \nonumber
b\times b &= c^2 \cong \mathbb{I} \\ \nonumber
f\times f &= \mathbb{I} \\ \nonumber
\sigma\times f &= \sigma \\ \nonumber
\sigma\times b &= \sigma \\ \nonumber
\sigma\times\sigma &= 1+f
\end{align}
and the topological spins are:
\begin{align}
\theta_b &= 1 \\ \nonumber
\theta_f &= -1 \\ \nonumber
\theta_\sigma &= e^{i\pi/8} 
\end{align}

\subsection{Moore-Read Quantum Hall State}
The Moore-Read state\cite{MooreRead} can be obtained from the KSL phase by placing $b$ in a $\nu=1/2$ bosonic-Laughlin quantum Hall phase rather instead of a Bose-Mott insulator.  This phase is characterized by the idealized wave-function:
\begin{align} \Psi_\text{MR} \sim  \prod_{i<j}(z_i-z_j)^2\text{Pf}\(\frac{1}{z_i-z_j}\) \end{align}
where $z_j= x_j+iy_j$ is the complexified coordinate of the $j^\text{th}$ electron.  The factors of $(z_i-z_j)^2$ stem from the $b$ sector, and the $\text{Pf}$ denotes the Pfaffian of the anti-symmetric matrix with entries $\frac{1}{z_i-z_j}$, which describes the BCS wave-function with $p+ip$-pairing\cite{ReadGreen}.

In this phase, the vison, $\sigma$ of the $f$-sector is bound to a $\pi$-flux of the Bosonic QH fluid which we denote $v$ (similarly, denote a $-\pi$ flux of the Bosonic QH fluid by $\bar{v}$).  A $\pi$-flux in a $\sigma_{H}=\frac{1}{2}$ system has charge $\frac{e}{4}$ and hence v has topological spin $e^{i\pi/8}$.  Denoting the non-Abelian vison/charge-$\frac{e}{4}$ vortex composite as $\sigma_v$, we have: $\theta_{\sigma_v} = e^{i\pi/4}$.  Since $\bar{v}$ is a $-\pi$-vortex bound to charge $-\frac{e}{4}$ it also contributes an extra $e^{i\pi/8}$ to the vison topological spin, indicating that the composite, $\sigma\times\bar{v}\equiv \sigma_v^{-1}$, has $\theta_{\sigma_v^{-1}} = \theta_{\sigma_v} = e^{i\pi/4}$.

\subsection{2D TR-Breaking Analog}
The MR state looks somewhat similar to the STO phase constructed above: there Ising non-Abelions attached to charged Abelian vortices.  However, unlike in the TI STO phase, $\sigma_v\times\sigma_v^{-1} = 1+f$ is charge-neutral.  More generally, since $\sigma_v$ and $\sigma_{\bar{v}}$ have opposite charge, and the same topological spin, it is hard to see how TR-invariance could be implemented in the MR phase, even at the surface of the STO.

We can cure this problem by introducing an extra counter-propagating anti-semion particle, $s$, with topological spin $\theta_s = e^{-i\pi/2}$ to the boson sector (in the parton language this corresponds to further fractionalizing $b\rightarrow b_1b_2$, with $b_1$ carrying charge $e$ in a bosonic $\nu=1/2$ QH phase, and $b_2$ a charge-neutral in a $\nu=-1/2$ bosonic QH phase).  Making the following identifications:
\begin{align}
\beta^{-1}\times\tau^2_+ &=s \nonumber\\
\tau_v &= \sigma_v \nonumber\\
\tvb &= \sigma_v\times s \nonumber\\
\tau_v^{-1} & = \sigma_v^{-1} \nonumber\\
\tvb^{-1} &= \sigma_v^{-1}\times s
\end{align}
we see that this 2D TR-breaking phase has the same topological order and charge assignments as the STO phase described above.  For brevity we denote the 2D TR-breaking phase: MR$\times$AS.

\section{Connection Between STO and Familiar Non-Fractionalized Surface Phases}
In the previous section, we have constructed an STO phase by quantum disordering the TRS surface superconductor state.  The fact that a TI can realize this topological order with both $U(1)_C$ and TR symmetries intact actually serves as a non-perturbative definition of the $U(1)_C\ltimes\mathbb{Z}_2^T$ fermion topological insulator.  To see this, we need to show that we can obtain all of the usual symmetry broken non-topologically ordered surface phases of the familiar fermion TI through a sequence of surface-phase transitions that do not affect the TI bulk.

\subsection{STO to TR-Symmetric Non-Abelian Surface SC}
Since we have constructed the STO phase from the TR-invariant surface SC, it is straightforward to recover the familiar surface SC.  We have already argued that the superconducting surface can be obtained from the STO phase by condensing $\beta$.  Here we provide some further details.

Since $\beta^2=b = \sqrt{\rho_b}e^{i\phi_b}$ we may write $\beta = (\rho_b)^{1/4}e^{i\phi_\beta}$.  Then $2\pi$ vortices of $\phi_\beta$ are $4\pi$ vortices of $\phi_b$, which are condensed in the STO phase.  In other words, the STO phase can be viewed as a Mott insulator of $\beta$.  Then, to recover the TRS surface superconductor from the STO phase, one can simply condense $\beta$.  Since $\beta$ has non-trivial mutual statistics with all other particles besides $f$, the particles $\tv,\tvb,\tau^2_\pm$ etc... will all be confined in the $\<\beta\>\neq 0$ phase.  However, these confined objects do not completely dissapear from the theory, rather they are bound to vortices of $\phi_\beta$ (which are now-gapped) to form composites that have trivial mutual statistics with the $\beta$-condensate.  

Since, $\beta$ has the same mutual statistics with $\tv$ as with a $\pi/2$-vortex of $\phi_\beta$ they are bound-together in the superconductor.  Since $\beta$ is charged a $+\pi/2$-vortex of $\phi_\beta$ has physical circulating charge current. and the $\tv$ object becomes the superconducting $\frac{hc}{2e}$ vortex (or, more generally, a $\frac{(4n+1)hc}{2e}$ vortex with $n\in\mathbb{Z}$).  Similarly, $\tvb$ becomes a $\frac{(4n-1)hc}{2e}$ vortex, and $\tau^2_\pm$ become a $\frac{nhc}{e}$ vortices (with $n$ odd).

\subsection{STO to 1/2-integer quantum Hall}
Next, we connect the STO to the $U(1)_C$ preserving but TR-breaking $\frac{1}{2}$-integer surface quantum Hall insulator (SQHI).  In the previous section, we showed that the topological order and charge assignments of the STO can be realized in strict 2D at the expense of breaking TRS.  The analogous TR breaking phase was equivalent to the Moore-Read QH phase with an extra neutral semion, denoted MR$\times$AS.  Importantly, the MR$\times$AS has $\sigma_H=\kappa_H=\frac{1}{2}$.  There is a closely related phase, which we denote $\overline{\text{MR}\times\text{AS}}$, obtained from MR$\times$AS by switching all of the particles of MR$\times$AS with their anti-particles, which has $\sigma_H=\kappa_H=-\frac{1}{2}$.

Starting with the STO phase of the TI, let us ``deposit" a layer of $\overline{\text{MR}\times\text{AS}}$ on the TI surface (or alternatively, imagine adjusting the interactions and other parameters of a layer of the bulk near the surface to drive that layer into the $\overline{\text{MR}\times \text{S}}$ phase).  Then, suppose we allow the $f$ particle of the $\overline{\text{MR}\times\text{AS}}$ to hybridize with  (i.e. tunnel into) the $f$ particle of the STO phase.  This confines each non-Abelian $\tau_v$ of the STO is bound to a similar non-Abelian  $\tv^{-1}$ of the deposited layer, thereby neutralizing the non-Abelian statistics of the composite object.   The resulting composites are all Abelian and have trivial self-statistics, and hence can be straightforwardly condensed (since TR symmetry is already broken).  In particular, if we condense the particles containing a $\tv$ of the STO layer and a $\tv^{-1}$ of the deposited layer, all other particles are trivially confined, and no excitations with fractional statistics remain.

We have thereby eliminated the surface-topological order, at the expense of breaking TR-symmetry on the surface.  What is the quantum Hall response of this non-fractionalized insulating state?

\begin{figure}[ttt]
\includegraphics[width=3in]{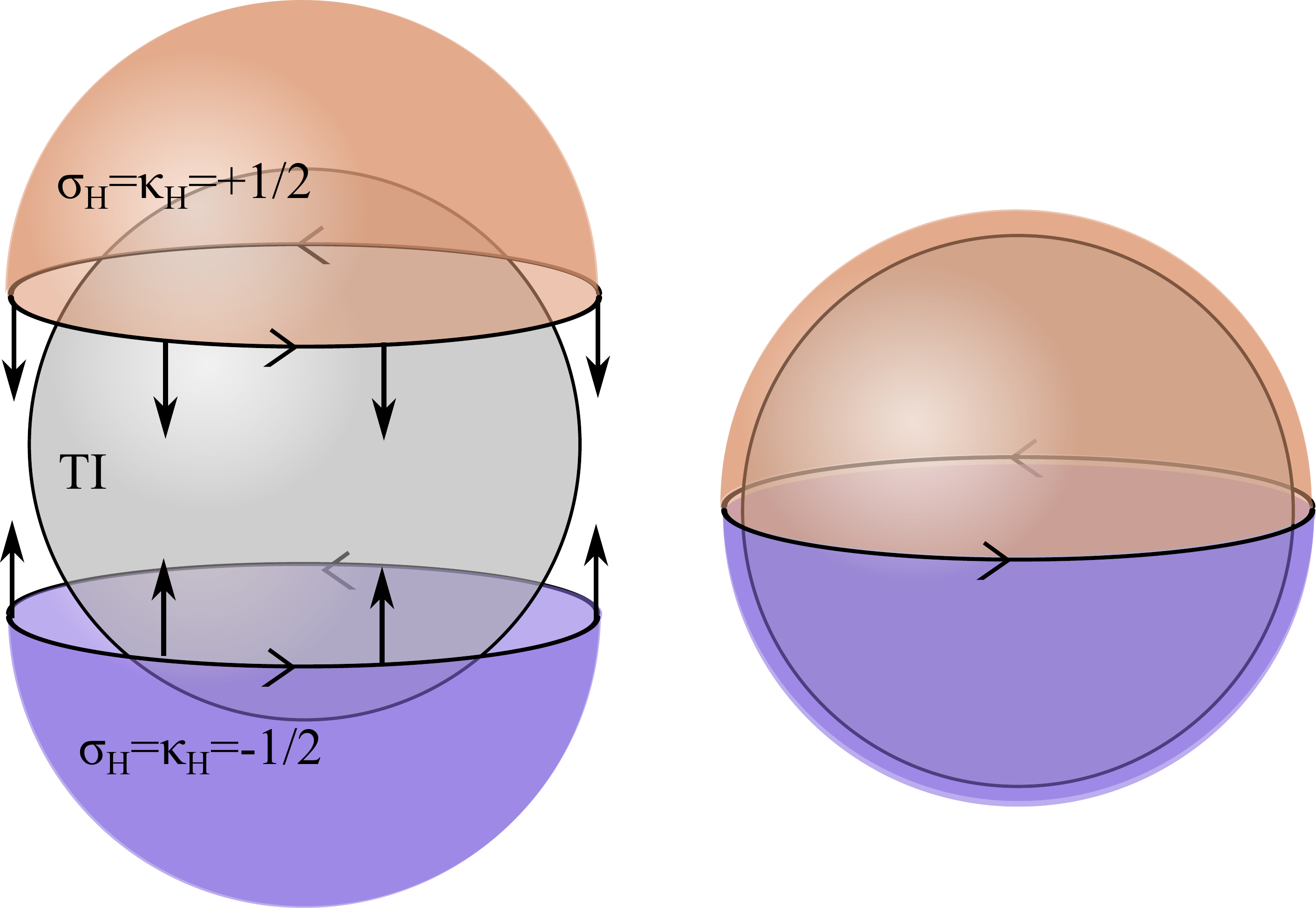}
\caption{The non-fractionalized TR-breaking quantum Hall insulator (QHI) with coating the TI surface with a 2D TR-breaking topologically ordered state with $\sigma_H=\kappa_H=\pm\frac{1}{2}$ (depicted in orange and purple respectively), as explained in the text.  The half-integer quantum Hall conductance can be seen by considering a domain between these two coatings as shown in the above figure for a spherical TI,}
\label{fig:TICoating}
\end{figure}

To answer this question we note that we could have equally well followed a time-reversed version of the above procedure, by depositing a different surface layer related to $\overline{\text{MR}\times\text{AS}}$ by TR, which we denote $\overline{\text{MR}^*\times \text{AS}^*}$ and has $\sigma_H=\kappa_H=\frac{1}{2}$.  Consider a spherical TI, depicted in Fig.~\ref{fig:TICoating}, and imagine depositing a layer of $\overline{\text{MR}\times\text{AS}}$ on the bottom hemisphere of the TI surface and a layer of $\overline{\text{MR}^*\times \text{AS}^*}$ on the top hemisphere.  The edges of the deposited 2D layers meet at the equator, and each contributes a chiral Majorana fermion, a co-propagating charged boson mode and a counter-propagating neutral boson mode.  The chiral Majorana fermions from the top and bottom hemisphere propagate in the same direction, and when coupled, combine into a complex (neutral) chiral fermion.  The combined edge has overall chirality with a single chiral charged mode, and hence has $\sigma_H=\kappa_H=1$.  This $\sigma_H$ and $\kappa_H$ is not effected by condensing $\tv$ composites in order to remove the topological order. 

This line of reasoning shows that, even after destroying the surface-topological order, the interface at the equator possesses a single 1D chiral charge fermion. The non-fractionalized phases that we have produced on the top and bottom hemisphere therefore differ by an electron $\nu=1$ quantum Hall layer.  Since these two phases are related by TR-symmetry, we must democratically assign them $\sigma_H=\kappa_H=\pm\frac{1}{2}$ respectively.  We have therefore succeeded in recovering the familiar non-fractionalized surface QH insulating phases from the STO phase.

\subsection{STO to Gapless Dirac Fermion Surface}
In the previous section, we showed how to obtain the surface QH insulator from the STO phase by breaking TR.  The resulting phase can have either $\sigma_H=\kappa_H = \pm \frac{1}{2}$.  From here, it is straightforward to produce the symmetry preserving gapless Dirac cone phase by proliferating domain walls between the $\sigma_H=\pm \frac{1}{2}$ surface phases.  Such domain walls carry a single chiral (complex) fermion, and it is well known (for example from network models\cite{AshvinSenthilSPT}) that their proliferation results in a single gapless Dirac cone.

\subsection{$\mathbb{Z}_2$ Nature of Surface Order}
It is well known that two copies of the ordinary electron topological insulator can be smoothly deformed into the trivial insulator without a bulk phase transition.  Therefore, as a final consistency check for the proposed STO, we demonstrate that two coupled STO phases can be deformed to a trivial insulator by surface phase-transitions that leave the bulk gap untouched.

Consider starting with two layers of the STO phase, labeled 1 and 2 respectively, coupled such that electrons can tunnel between them: $\<c_1^\dagger c_2\>$, $\<c_2^\dagger c_1\>\neq 0$.  It is straightforward to check that the following set of composite particles are charge-neutral self-bosons with trivial mutual-statistics, which can be simultaneously condensed without breaking either $U(1)_C$ or TRS:
\begin{align} \{\beta_1^\dagger\beta_2,~~\tau_{v1}\tau_{v2}\beta^\dagger,~~ \tau_{\bar{v}1}\tau_{\bar{v}2}\beta^\dagger,~~\text{and h.c.'s}\}
\label{eq:ParticleList} \end{align}
with h.c.'s indicating that all operators related by Hermitian conjugation to those listed are also condensed. In order to preserve TRS, we must condense TR conjugate particles with equal amplitude: $\<\tau_{v1}\tau_{v2}\beta^\dagger\>= \<\tau_{\bar{v}1}\tau_{\bar{v}2}\beta^\dagger\>\neq 0$.

It is also straightforward to verify that after condensing these objects,  all non-trivial particles in the theory are either confined or condensed, and there are no fractionalized excitations.  In particular, $f$ and $\beta^2$ both have mutual $(-1)$ statistics with the condensed $\tau_{v1}\tau_{v2}\beta^\dagger$ particles, and are confined together to form the physical electron: $c=\beta^2 f$.  The resulting phase has only gapped, physical electron excitations, $c$, and hence is a trivial band-insulator.  Therefore, we have verified that the bulk phase described by our proposed STO indeed has a $\mathbb{Z}_2$ group structure (i.e. that combining two copies of our phase produces a trivial phase) as required for the electron TI.

This set of particles in Eq.~\ref{eq:ParticleList} has a natural physical interpretation: starting with two coupled layers of the TRS surface-SC phase, we know that we can obtain a trivial bulk insulator by condensing the (now Abelian) $\pm\frac{hc}{2e}$ vortices, which now occur in the same location in both layers due to the interlayer tunneling.  The set of particles condensed here to trivialize the double-layer STO phase are simply the descendants of these vortices.

\section{Discussion}
We have shown that, in addition to the familiar gapless Dirac surface state, and gapped symmetry-broken states, the electronic topological insulator (TI) can support a gapped and fully symmetric phase with surface topological order (STO).  This STO phase provides a complete, non-perturbative definition of the electron TI.  Like STO phases of analogous bosonic TIs, the electron TI STO phase has the same topological-order as a 2D phase, but with symmetry implemented in a way that is not allowed in strict 2D.  

For boson TIs, the lens of STO provides a useful perspective into 3+1D strongly correlated boson TIs as well as 2+1D gauge theories\cite{ChongSenthil}.  The hope is then that understanding of the electron TI STO will enable similar progress for strongly-correlated electronic phases.  An essential component for boson TIs was a systematic understanding of symmetry implementation for strictly-2D Abelian bosonic systems\cite{LuVishwanath,LevinStern}. One potentially complicating factor in adapting this approach to fermions is that the electron TI STO is inherently non-Abelian. Consequently an important outstanding task for making progress along these lines is to develop a systematic understanding of symmetry implementation in 2D non-Abelian theories.  These theories are not amenable to the simple K-matrix methods that have so successfully utilized for boson systems\cite{LuVishwanath,LevinStern}.  However, methods of similar spirit based on using the bulk-boundary correspondence to reduce the problem to symmetry implementation in 1+1D conformal field theories of the edge may still prove fruitful.  Such a pursuit would go far beyond the scope of the present paper and is left as a challenge for future work.

Using a different method, based on Walker-Wang type models, X. Chen, L. Fidkowski, and A. Vishwanath have also constructed a candidate STO phase for a $\Theta=\pi$ electron TI\cite{BerkeleySTO}.  The relationship between this STO and the one described above is not completely clear, however, in light of the general arguments of Ref.~\onlinecite{fSPTClassification} this phase can at most differ from the conventional eTI by an SPT phase of neutral bosons.

\textit{Acknowledgements - } We thank A. Vishwanath and X. Chen for inspiring discussions, and for sharing and discussing their unpublished work. TS thanks the hospitality of Harvard University where this work was partially done.  ACP was supported by a an interdisciplinary quantum information science and engineering fellowship (NSF Grant No. DGE-0801525).  TS and CW were supported by Department of Energy DESC-8739-ER46872, and partially supported by the Simons Foundation by award number 229736.

During the completion of this manuscript, we became aware of a similar work by M. Metlitski, C.L. Kane, and M.P.A. Fisher which obtains the same STO\cite{Metlitski-eTI}, and a related work by P. Bonderson, C. Nayak, X.L. Qi\cite{QSTO} which obtains the same STO as in Ref.\onlinecite{BerkeleySTO}.


\begin{thebibliography}{99}
\bibitem{Endnote:VortexStatistics}
To precisely define vortex statistics, it is necessary to consider a gauged $U(1)_C$ symmetry (as is the case for real electrons coupled to the physical electromagnetic field, which has very weak fluctuations).
\end{thebibliography}
\end{document}